\PassOptionsToPackage{unicode}{hyperref}
\PassOptionsToPackage{hyphens}{url}
\PassOptionsToPackage{dvipsnames,svgnames,x11names}{xcolor}
\documentclass[
]{article}
\usepackage{amsmath,amssymb, authblk}
\usepackage{caption}
\usepackage{subcaption}
\usepackage{graphicx}
\usepackage{lmodern}
\usepackage{geometry}
\usepackage{iftex}
\usepackage{natbib} 
\ifPDFTeX
  \usepackage[T1]{fontenc}
  \usepackage[utf8]{inputenc}
  \usepackage{textcomp} % provide euro and other symbols
\else % if luatex or xetex
  \usepackage{unicode-math}
  \defaultfontfeatures{Scale=MatchLowercase}
  \defaultfontfeatures[\rmfamily]{Ligatures=TeX,Scale=1}
\fi
\IfFileExists{upquote.sty}{\usepackage{upquote}}{}
\IfFileExists{microtype.sty}{% use microtype if available
  \usepackage[]{microtype}
  \UseMicrotypeSet[protrusion]{basicmath} % disable protrusion for tt fonts
}{}
\makeatletter
\@ifundefined{KOMAClassName}{% if non-KOMA class
  \IfFileExists{parskip.sty}{%
    \usepackage{parskip}
  }{% else
    \setlength{\parindent}{0pt}
    \setlength{\parskip}{6pt plus 2pt minus 1pt}}
}{% if KOMA class
  \KOMAoptions{parskip=half}}
\makeatother
\usepackage{xcolor}
\usepackage{color}
\usepackage{fancyvrb}

\DefineVerbatimEnvironment{Highlighting}{Verbatim}{commandchars=\\\{\}}
\usepackage{graphicx}
\makeatletter
\def\maxwidth{\ifdim\Gin@nat@width>\linewidth\linewidth\else\Gin@nat@width\fi}
\def\maxheight{\ifdim\Gin@nat@height>\textheight\textheight\else\Gin@nat@height\fi}
\makeatother
\setkeys{Gin}{width=\maxwidth,height=\maxheight,keepaspectratio}
\makeatletter
\def\fps@figure{htbp}
\makeatother
\makeatletter
 \let\@cite@ofmt\@firstofone
 \def\@biblabel#1{}
 \def\@cite#1#2{{#1\if@tempswa , #2\fi}}
\makeatother
\newlength{\cslhangindent}
\newlength{\csllabelwidth}
 {\begin{list}{}{%
  \setlength{\itemindent}{0pt}
  \setlength{\leftmargin}{0pt}
  \setlength{\parsep}{0pt}
  \ifodd #1
   \setlength{\leftmargin}{\cslhangindent}
   \setlength{\itemindent}{-1\cslhangindent}
  \fi
  \setlength{\itemsep}{#2\baselineskip}}}
 {\end{list}}
\usepackage{calc}

\ifLuaTeX
\usepackage[bidi=basic]{babel}
\else
\usepackage[bidi=default]{babel}
\fi
\babelprovide[main,import]{american}
\def\languageshorthands#1{}
\ifLuaTeX
  \usepackage{selnolig}  % disable illegal ligatures
\fi
\IfFileExists{bookmark.sty}{\usepackage{bookmark}}{\usepackage{hyperref}}
\IfFileExists{xurl.sty}{\usepackage{xurl}}{} % add URL line breaks if available
\hypersetup{
  pdftitle={bayes\_spec: A Bayesian Spectral Line Modeling Framework for
Astrophysics},
  pdfauthor={Trey V. Wenger},
  pdflang={en-US},
  colorlinks=true,
  linkcolor={Maroon},
  filecolor={Maroon},
  citecolor={Blue},
  urlcolor={Blue},
  pdfcreator={LaTeX via pandoc}}

\title{CompactObject: An open-source Python package for full-scope neutron star equation of state inference}

\usepackage{orcidlink}

\author[1%
  ]{Chun Huang%
    \,\orcidlink{0000-0001-6406-1003}\,%
    \thanks{chun.h@wustl.edu}
    }
\author[2%
  ]{Tuhin Malik%
    \,\orcidlink{0000-0003-2633-5821}\,%
    }
\author[2%
  ]{João Cartaxo%
    \,\orcidlink{0009-0001-7105-8272}\,%
    }
\author[1%
  ]{Shashwat Sourav%
    \,\orcidlink{0000-0002-0169-4003}\,%
    }
\author[3%
  ]{Wenli Yuan%
    \,\orcidlink{0000-0003-2771-759X}\,%
    }
\author[4%
  ]{Tianzhe Zhou%
    \,\orcidlink{0009-0000-8504-9134}\,%
    }
\author[5%
  ]{Xuezhi Liu%
   \,\orcidlink{0009-0008-3286-7254}\,%
    }
\author[1%
  ]{John Groger%
    \,\orcidlink{0000-0002-7054-9053}\,%
    }
\author[6%
   ]{Xieyuan Dong%
      \,\orcidlink{0009-0001-6228-1317}\,%
     }
\author[1%
  ]{Nicole Osborn%
    \,\orcidlink{}\,%
    }
\author[1%
  ]{Nathan Whitsett%
    \,\orcidlink{0000-0002-7960-8064}\,%
    }
\author[1%
  ]{Zhiheng Wang%
    \,\orcidlink{0009-0000-5088-6207}\,%
    }
\author[2%
  ]{Constan\c{c}a Provid\^{e}ncia%
    \,\orcidlink{0000-0001-6464-8023}\,%
    }
\author[7%
  ]{Micaela Oertel%
    \,\orcidlink{0000-0002-1884-8654}\,%
    }
    \author[1%
  ]{Alexander Y. Chen%
    \,\orcidlink{0000-0002-4738-1168}\,%
    }
\author[8,9,10%
  ]{Laura Tolos%
    \,\orcidlink{0000-0003-2304-7496}\,%
    }
\author[11%
  ]{Anna Watts%
    \,\orcidlink{0000-0002-1009-2354}\,%
    }

\affil[1]{Physics Department and McDonnell Center for the Space Sciences, Washington University in St. Louis, MO 63130, USA%
  }
\affil[2]{CFisUC, Department of Physics, University of Coimbra, 3004-516 Coimbra, Portugal%
  }
\affil[3]{School of Physics and State Key Laboratory of Nuclear Physics and Technology, Peking University, Beijing 100871, China%
  }
\affil[4]{Department of Physics, Tsinghua University, Beijing 100084, China%
  }
\affil[5]{Physics Department, Central China Normal University, Luoyu Road, 430030, Wuhan, China%
  }
 \affil[6]{School of Physics, Nankai University, Tianjin 300071, China%
   }
\affil[7]{Observatoire astronomique de Strasbourg, CNRS, Université de Strasbourg, 11 rue de l'Université, 67000 Strasbourg, France
  }
\affil[8]{Institute of Space Sciences (ICE, CSIC), Campus UAB, Carrer de Can Magrans, 08193, Barcelona, Spain%
  }
\affil[9]{Institut d'Estudis Espacials de Catalunya (IEEC), 08860 Castelldefels (Barcelona), Spain%
  }
\affil[10]{Frankfurt Institute for Advanced Studies, Ruth-Moufang-Str. 1, 60438 Frankfurt am Main, Germany%
  }
\affil[11]{Anton Pannekoek Institute for Astronomy, University of Amsterdam, Science Park 904, 1090 GE Amsterdam, the Netherlands}
\date{\today}

\begin{document}
\maketitle

\section{Summary}\label{summary}

The CompactObject package is an open-source software framework developed to constrain the neutron star equation of state (EOS) through Bayesian statistical inference. It integrates astrophysical observational constraints from X-ray timing, gravitational wave events, and radio measurements, as well as nuclear experimental constraints derived from perturbative Quantum Chromodynamics (pQCD) and Chiral Effective Field Theory ($\chi$EFT). The package supports a diverse range of EOS models, including meta-model like and several physics-motivated EOS models. It comprises three independent components: an EOS generator module that currently provides seven EOS choices, a Tolman–Oppenheimer–Volkoff (TOV) equation solver, that allows the determination of the  Mass Radius and Tidal deformability as observables, and a comprehensive Bayesian inference workflow module, including a complete pipeline for implementing EOS  Bayesian inference. Each component can be used independently in different scientific research contexts, such as nuclear physics and astrophysics.  In addition, CompactObject is designed to work in synergy with existing software such as  \href{https://compose.obspm.fr}{CompOSE}, allowing the use of the CompOSE EOS database~\cite{Typel:2013rza,CompOSECoreTeam:2022ddl} to extend the EOS options available.

\section{Statement of need}\label{statement-of-need}

Understanding the equation of state (EOS) of neutron stars is important for understanding the fundamental physics governing ultra-dense matter. Neutron stars, with their core densities exceeding several times nuclear saturation density, have a crucial role to play in studying nuclear interactions under extreme conditions. However, inferring the EOS from observational and experimental data has significant challenges due to the complex interplay of astrophysical phenomena and nuclear physics. Many of these studies such as \cite{Raaijmakers2023} focus on EOS meta-models (which may be parameterized or non-parameterized) that attempt to span all reasonable mass-radius parameter space, rather than being driven by microphysics. In contrast, CompactObject achieves high accuracy and rapid computation for a family of physics-motivated EOSs, thereby enabling researchers to perform inferences based on physically motivated models and apply nuclear physics-related constraints derived from nuclear experiments. %Existing tools often lack the integration of diverse datasets required for comprehensive Bayesian analysis, limiting the precision and scope of EOS constraints.

\par CompactObject appears as a viable solution to these challenges by providing an open-source, robust platform designed for Bayesian inference on neutron star EOS constraints. Its comprehensive workflow integrates a wide range of EOSs, including not only physical microscopic models  and meta-models of neutron star EOSs but also %strange star and 
quark star EOSs, i.e., strange stars \cite{Bodmer:1971,Witten:1984}, nonstrange stars \cite{Holdom:2018} or strangeon star EOSs \citep{2003ApJ...596L..59X}, which have been proposed to explain the nature of these compact objects, enabling a detailed exploration of dense matter physics. The package's user-friendly interface and modular architecture facilitates a easy adoption and extension, allowing researchers to customize analyses and incorporate new EOSs as they become available. Furthermore, thorough documentation ensures that both novice and experienced users can effectively utilize the tool, promoting widespread accessibility and collaborative advancement in the field. By addressing the need for an integrated, flexible, and well-documented framework, CompactObject enhances the capability of nuclear astrophysicists to derive precise EOS constraints. 
%This, in turn, improves our understanding of fundamental nuclear interactions, the behavior of matter at supra-nuclear densities, and the diverse phenomena associated with neutron stars. 
\section{The CompactObject Package and scientific use}\label{usage}
CompactObject is an open-source software package designed to apply astrophysical and nuclear physics constraints to EOS parameters. Currently, the available EOS options include polytropic EOSs and speed of sound model EOSs, both of which are meta-models. Additionally, the package supports physics-motivated models such as the Relativistic Mean Field (RMF) theory \citep{Todd-Rutel:2005yzo,Tolos_2016,Tolos_Centelles_Ramos_2017} and its density-dependent variant \citep{Typel:1999yq,Hempel_2010,Char_2014,Char:2023fue}. We have  integrated features for users to define the density dependent variant form by themselves, and which span most of the possibility of this family of models. Beyond neutron star EOS models, CompactObject also includes 
the widely used MIT bag model \citep{PhysRevD.9.3471}  and a strangeon matter EOS model \citep{2003ApJ...596L..59X} for quark stars, which are analytical and convenient for statistical analysis. Meanwhile, this package also includes the more advanced quark matter EOS derived from the Nambu-Jona-Lasinio (NJL) model \cite{Klevansky:1992,Buballa:2005}, which is motivated by the basic symmetries of QCD.
The package integrates various likelihood constraints, including routines for simulating mass-radius measurements from X-ray timing observations and analyzing mass-radius likelihoods from actual observational data. It also incorporates constraints from radio timing observations, gravitational wave observations related to tidal deformability, and nuclear physics constraints derived from saturation properties, pQCD \cite{Gorda:2022jvk}, and $\chi$EFT \cite{Hebeler:2013nza, Huth:2021bsp}. 

Furthermore, CompactObject includes routines for EOS analysis that output additional properties of neutron stars, such as proton fraction and the number densities of different particles within the star. Other than these, CompactObject synergizes with the CompOSE database, allowing users to derive observational evidence directly into existing EOS models. The nested sampling pipeline implemented in CompactObject is based on UltraNest, providing a computational framework to extract Bayesian evidence for each integrated EOS model. For the inference pipeline, the package offers two sampling algorithm options: UltraNest (nested sampling) and emcee (Markov Chain Monte Carlo sampling) 

CompactObject has been utilized to derive constraints on nucleonic RMF models \citep{Huang:2023grj} and hyperonic RMF models \citep{Huang:2024rvj}. Ongoing projects include constraining the strangeon star EOS \cite{Yuan2024} and exploring phase transitions and twin stars. Additionally, various nuclear and astrophysical constraints on RMF model with density-dependent couplings and non-linear mesonic terms have been  implemented in  \cite{Malik:2022zol,Malik:2023mnx,Providencia:2023rxc}.

The released version of CompactObject is readily accessible through its GitHub repository \citep{EoS_inference} under the MIT license and is archived on Zenodo repository \citep{COZenodo}. Comprehensive documentation and the complete workflow for implementing EOS inference are available in the GitHub repository. Future plans for CompactObject include expanding the range of available EOS options and conducting a detailed survey of existing EOS models to perform cross-comparisons of Bayesian evidence using current observational and experimental constraints.

\texttt{Software:} Python language \citep{10.1109/MCSE.2007.58}, Numpy \citep{van_der_Walt_2011}, MPI for Python \citep{DALCIN2008655}, Numba \citep{numba}, NumbaMinpack \citep{wogan_NumbaMinpack}, Matplotlib \citep{Hunter:2007}, Jupyter \citep{2016ppap.book...87K}, UltraNest \citep{2021JOSS....6.3001B}, emcee \citep{Foreman_Mackey_2013}, SciPy \citep{2020SciPy-NMeth}, Seaborn \citep{Waskom2021}, corner.py \citep{corner}

\section{Acknowledgments}
H.C., S.S., O.N., W.N. and J.G. acknowledge support from the Arts \& Sciences Fellowship of Washington University in St Louis. H.C. also acknowledges support from NASA grant 80NSSC24K1095. 
C.P., T. M. and J. C.  received support from Fundação para a Ciência e a Tecnologia (FCT), I.P., Portugal, under the  projects UIDB/04564/2020 (doi:10.54499/UIDB/04564/2020), UIDP/04564/2020 (doi:10.54499/UIDP/04564/2020), and 2022.06460.PTDC (doi:10.54499/2022.06460.PTDC).
L.T. acknowledges support from CEX2020-001058-M (Unidad de Excelencia ``Mar\'{\i}a de Maeztu") and PID2022-139427NB-I00 financed by the Spanish MCIN/AEI/10.13039/501100011033/FEDER,UE as well as from the Generalitat de Catalunya under contract 2021 SGR 171,  from the Generalitat Valenciana under contract CIPROM/2023/59 and by the CRC-TR 211 'Strong-interaction matter under extreme conditions'- project Nr. 315477589 - TRR 211. A.L.W. acknowledges support from ERC Consolidator Grant No.~865768 AEONS. A.C. acknowlege the support from NSF grants DMS-2235457 and AST-2308111. M. O. acknowledges financial support from the Agence Nationale de la recherche (ANR) under contract number ANR-22-CE31-0001-01.

\bibliography{joss}{}
\bibliographystyle{aasjournal}% Produces the bibliography via BibTeX.

\end{document}